\documentclass[aps,twocolumn]{revtex4}
\usepackage{amssymb}

\usepackage{amsmath}
\usepackage{graphicx}

\input{tcilatex}

\begin{document}

\title{Thermoelectric efficiency in the space-charge-limited transport
regime in semiconductors}
\author{Fran\c{c}ois L\'{e}onard}
\email{fleonar@sandia.gov}
\date{\today }

\begin{abstract}
The thermoelectric efficiency of semiconductors is usually considered in the
ohmic electronic transport regime, which is achieved through high doping.
Here we consider the opposite regime of low doping where the current-voltage
characteristics are nonlinear and dominated by space-charge-limited
transport. We show that in this regime, the thermoelectric efficiency can be
described by a single figure of merit, in analogy with the ohmic case.
Efficiencies for bulk, thin film, and nanowire materials are discussed, and
it is proposed that nanowires are the most promising to take advantage of
space-charge-limited transport for thermoelectrics.
\end{abstract}

\maketitle

\address{Sandia National Laboratories, Livermore, California 94551}

\section{\protect\bigskip Introduction}

Most implementations of thermoelectric devices utilize highly doped
semiconductors with linear (ohmic) current-voltage characteristics. In such
materials, the thermoelectric efficiency is determined by the figure of
merit $zT=S^{2}\sigma T/\kappa $ where $S$ is the Seebeck coefficient, $%
\sigma $ the electrical conductivity, $T$ the temperature, and $\kappa $ the
thermal conductivity. The $zT$ factor has governed much of the research in
developing approaches to improve thermoelectric efficiency; however, the
intimate relationship between transport quantities places constraints on the
possible paths towards efficiency improvements. Thus, new approaches where
the thermoelectric efficiency does not depend on $zT$ would open new routes
for scientific exploration.

An example of such an approach is thermionic cooling and power generation,
which was originally considered for injection into vacuum\cite{mahan}, and
then into semiconductors\cite{shakouri}. These thermionic devices are often
considered different from thermoelectric devices due to the presence of an
injection barrier and because the channel length is shorter than the
electronic mean-free path, giving ballistic electron transport\cite%
{nolas,ulrich,mahan2}. However, as the channel length increases beyond the
mean-free path, thermionic devices become equivalent to thermoelectric
devices\cite{mahan3,zeng}, which has led to new proposals to engineer
multilayer semiconductor materials\cite{odwyer}.

In this manuscript, we explore an alternative approach that does not rely on
ohmic transport or thermionic emission, but instead exploits
space-charge-limited (SCL) electronic transport in a semiconductor. We
derive an expression for thermoelectric efficiency in this regime, and show
that it depends on a new dimensionless figure of merit that replaces $zT$.
Furthermore, we apply the theory to bulk, thin film, and nanowire materials,
and conclude that nanowires are the most promising to harness SCL transport
for thermoelectrics.

\begin{figure}[h]
\centering \includegraphics[scale=0.4,clip=true]{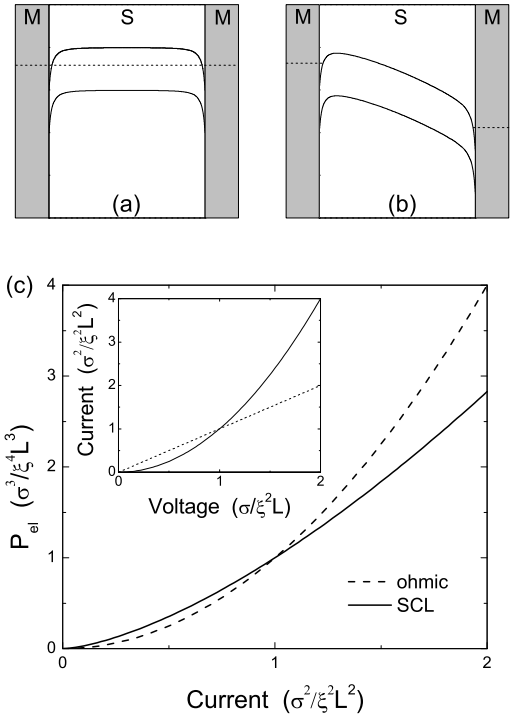}
\caption{Panels (a) and (b) show band diagrams for a
metal-semiconductor-metal system when the semiconductor has low doping.
Figure (a) is at equilibrium while figure (b) is under an applied voltage.
Solid lines are the valence and conduction band edges and dashed lines are
the Fermi levels. (c) Power loss due to Joule heating plotted for ohmic
transport and for SCL transport in dimensionless units. The inset shows the
current-voltage characteristics for these two transport regimes.}
\end{figure}

The term "space-charge-limited" originated from the consideration of charge
injection into a region where electronic transport is ballistic. In that
case, space-charge effects are detrimental because they create an additional
barrier for injection compared with the reference space-charge-free system%
\cite{lampert}. This situation is relevant to {\it thermionic} devices where
space-charge effects have been shown to reduce efficiency\cite{mahan}. The
issue of space-charge is different if one considers semiconductor-based {\it %
thermoelectric} devices. There, the reference system is a highly-doped
semiconductor with two ohmic contacts. In that case, space-charge effects
are minimal because the high doping immediately screens the injected charge.
However, if the doping in the semiconductor is lowered, space charge effects
become dominant, and the current becomes larger than the ohmic current
because electronic transport is no longer determined by the free-carrier
relaxation time, but by the shorter carrier transit time\cite{lampert}. This
situation, which has been observed experimentally in a broad range of
materials, is the one that we consider in this manuscript.

Figure 1 shows the system under consideration: a low-doped semiconductor
between two ohmic contacts. Band diagrams are shown in Figs 1a,b for the
case relevant to electron injection: the metal Fermi level contacts the
semiconductor in the conduction band creating a large density of carriers in
the near-interface region. Band-bending away from the metal/semiconductor
interface creates a barrier for electron injection at zero bias. Under bias,
a maximum in the potential is created near the injecting electrode, defining
the so-called "virtual cathode"\cite{rose}. Modeling has demonstrated that
in the presence of this virtual cathode, SCL transport dominates as soon as
the applied bias $V$ exceeds a few $kT$\cite{grinberg1}. This leads to a
current-voltage relationship $J\propto V^{2}$ as discussed further below.
This relationship differs from the exponential $J-V$ characteristics usually
considered for thermionic systems\cite{mahan}; the transitions between the
thermionic and SCL regimes and the role of Schottky barriers has been
studied in detail\cite{davids,chandra}, demonstrating that SCL dominates
even when small Schottky barriers are present.

\section{Thermoelectric efficiency}

To be specific, we consider the thermoelectric power generation efficiency
of the system of Fig. 1, but the approach should apply to cooling as well.
The efficiency of a thermoelectric material for power generation is given by%
\cite{snyder}%
\begin{equation}
\eta =\frac{P_{th}-P_{el}}{Q}=\frac{J\int_{T_{c}}^{T_{h}}S(x)dT-J%
\int_{0}^{L}\nabla v(x)dx}{JT_{h}S_{h}+\kappa _{h}\nabla T_{h}}
\label{efficiency}
\end{equation}%
where $P_{th}$ is the thermoelectric power generated, $P_{el}$ is the
electrical power dissipated, and $Q$ is the thermal power supplied. Here $%
T_{c}$ and $T_{h}$ are the cold and hot side temperatures and $v(x)$ is the
local potential.

Usually, one considers a highly doped semiconductor of length $L$ between
ohmic contacts. For a bias voltage $V$, there is a linear drop of the
potential along the channel due to the free-carrier screening; this leads to 
$J=\sigma V/L$, giving $P_{el}=J^{2}L/\sigma $ and the dependence of the
efficiency on $zT$\cite{snyder}. For SCL transport, the transport equations
need to be solved self-consistently with the Poisson equation due to the
unscreened space-charge\cite{grinberg}; in the simplest model with diffusive
electronic transport\cite{grinberg}, this gives an electrostatic potential
that depends on position as 
\begin{equation}
v(x)=\left( x/L\right) ^{3/2}V.
\end{equation}%
This nonlinear spatial dependence of the potential leads to the
current-voltage relationship \cite{grinberg} 
\begin{equation}
J=\xi ^{2}V^{2}.
\end{equation}%
Thus, in the SCL regime, the current is no longer linear in $V$, but is
quadratic instead. The proportionality constant $\xi $ depends on the
dimensionality of the material, and to some extent on the shape of the
electrodes, as will be discussed further below.

The quadratic form of the $J-V$ behavior implies that $P_{el}=J^{3/2}/\xi $;
the interest in this transport regime for thermoelectrics is illustrated in
Fig. 1c, where $P_{el}$ for the ohmic and SCL regimes are compared. There, $%
P_{el}$ in the SCL regime is below that of the ohmic regime at sufficiently
high voltage, suggesting a potentially new regime of operation for
thermoelectrics. (The cross-over between ohmic and SCL transport occurs at $%
V_{c}=\sigma /\xi ^{2}L$, as shown in the Fig. 1c inset. This sets a
condition for when ohmic contributions should be negligible compared to SCL
transport.)

To derive the thermoelectric efficiency in this regime, we utilize the
approach described in Ref. \cite{snyder}. From the above equations we obtain
for the local efficiency

\begin{equation}
d\eta (x)=\frac{dT}{T}\frac{S-\frac{3}{2}\frac{J^{1/2}}{\xi \Delta T}\left( 
\frac{x}{L}\right) ^{1/2}}{S+\kappa \nabla T/TJ}=\frac{dT}{T}\eta _{r}\left(
x,T\right) ,
\end{equation}%
where we used $dP_{el}=Jdv(x)$, and where the prefactor $dT/T$ indicates
that the efficiency is limited by the Carnot efficiency. The temperature $T$
depends on position through the heat equation\cite{leonard}%
\begin{equation}
\frac{d\left( \kappa \nabla T\right) }{dx}=-T\frac{dS}{dT}J\nabla T-\frac{3}{%
2}\frac{J^{3/2}}{\xi L}\left( \frac{x}{L}\right) ^{1/2}.
\end{equation}%
The first term on the right hand side is the usual Thomson effect, while the
second term is the local Joule heating in the material $dP_{el}$. This
equation applies for SCL transport because both the electronic and phononic
transport are diffusive, and thus both the electronic and phononic thermal
conductivities are well defined (this has been demonstrated for SCL
transport in nanowires\cite{katzenmeyer}).

The total efficiency of a segment is obtained from\cite{harman}

\begin{equation}
\eta =1-\exp \left[ -\int_{T_{c}}^{T_{h}}\frac{\eta _{r}\left( x,T\right) }{T%
}dT\right] .
\end{equation}%
Noticing that $d\ln \left( ST+\kappa \nabla T/J\right) /dT=[S+d(\kappa
\nabla T/J)/dT]/[ST+\kappa \nabla T/J]$ and assuming that the coefficients
are independent of temperature and position we arrive at

\begin{equation}
\eta =1-\frac{S_{c}T_{c}+\kappa \nabla T_{c}/J}{S_{h}T_{h}+\kappa \nabla
T_{h}/J}.
\end{equation}%
Thus, the efficiency depends only on quantities evaluated at the cold and
hot sides\cite{snyder}. To obtain these quantities we integrate the heat
equation to get 
\begin{eqnarray}
\kappa \nabla T_{c}/J &=&\frac{\kappa }{J}\left( \frac{\Delta T}{L}+\frac{2}{%
5}\frac{J^{3/2}}{\xi \kappa }\right)  \nonumber \\
\kappa \nabla T_{h}/J &=&\frac{\kappa }{J}\left( \frac{\Delta T}{L}-\frac{3}{%
5}\frac{J^{3/2}}{\xi \kappa }\right)
\end{eqnarray}%
which leads to

\begin{equation}
\eta =\frac{\Delta T}{T_{h}}\frac{y^{2}\left( 1-y\right) }{y^{2}+\frac{1}{g%
\overline{T}}\frac{T_{h}}{\Delta T}-\frac{3}{5}\frac{\Delta T}{T_{h}}y^{3}}
\label{effy}
\end{equation}%
where 
\begin{equation}
y=\left( \frac{J}{S^{2}\xi ^{2}\left( \Delta T\right) ^{2}}\right) ^{1/2}
\label{y}
\end{equation}%
and 
\begin{equation}
g\overline{T}=\frac{S^{3}\xi ^{2}LT_{h}^{2}}{\kappa }.
\end{equation}%
The variable $y$ is a dimensionless quantity equal to the ratio between the
generated voltage and the maximum thermoelectric voltage that can be
developed across the material. $g\overline{T}$ is a dimensionless parameter
that replaces the $zT$ factor from conventional ohmic materials. (Note that
the average temperature $\overline{T}$ does not directly appear in $g%
\overline{T}$. However, we use the notation $\overline{T}$ as a reminder
that, as a first approximation, the quantities that enter $g\overline{T}$
should be evaluated at the average temperature across the material.)

\begin{figure}[h]
\centering \includegraphics[scale=0.5,clip=true]{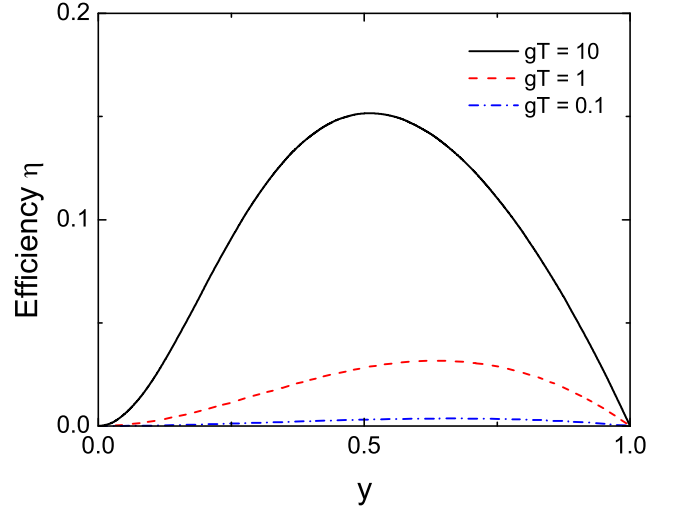}
\caption{Thermoelectric efficiency as a function of variable $y$ for three
values of the $g\overline{T}$ factor.}
\end{figure}

Figure 2 shows the dependence of the efficiency on $y$ from Eq. $\left( \ref%
{effy}\right) $, indicating that positive efficiency is achieved in the
range $0<y<1$, with a maximum attained at an intermediate value. This
behavior is similar to that obtained for conventional ohmic losses\cite%
{snyder}, but with a different functional dependence.

The maximum efficiency is obtained by maximizing $\eta $ with respect to $y$
to obtain%
\begin{equation}
\frac{g\overline{T}}{2}\frac{\Delta T}{T_{h}}\left( 1-\frac{3}{5}\frac{%
\Delta T}{T_{h}}\right) y^{3}+\frac{3}{2}y-1=0.
\end{equation}%
This equation provides the optimal $y\left( g\overline{T},\Delta
T/T_{h}\right) $ that maximizes the efficiency:%
\begin{equation}
y^{\ast }=\frac{1}{2\epsilon }\delta -\frac{1}{\delta }
\end{equation}%
where $\delta =\sqrt{\epsilon }\left[ 4\sqrt{\epsilon }+2\sqrt{2}\sqrt{%
1+2\epsilon }\right] ^{1/3}$and $\epsilon =\frac{g\overline{T}}{2}\frac{%
\Delta T}{T_{h}}\left( 1-\frac{3}{5}\frac{\Delta T}{T_{h}}\right) $. At this
optimal $y$ the efficiency is given by%
\begin{equation}
\eta =\frac{\Delta T}{T_{h}}\frac{1-\frac{3}{2}y^{\ast }}{1-\frac{9}{10}%
\frac{\Delta T}{T_{h}}y^{\ast }}.
\end{equation}%
Thus, the efficiency only depends on $g\overline{T}$ and $\Delta T/T_{h}$,
much like ohmic materials where the dependence is on $zT$ and $\Delta
T/T_{h} $.

Figure 3a shows the calculated efficiency as a function of $g\overline{T}$
for $\Delta T/T_{h}=1/2$, indicating low efficiency at small $g\overline{T}$
and efficiencies approaching the Carnot efficiency at large $g\overline{T}$.
The two limiting cases of small and large $g\overline{T}$ can be obtained as 
\begin{equation}
\eta =\left\{ 
\begin{array}{l}
\frac{\Delta T}{T_{h}}\left[ 1-3\left( \frac{1}{2}-\frac{3}{10}\frac{\Delta T%
}{T_{h}}\right) ^{2/3}\left( \frac{\Delta T}{T_{h}}g\overline{T}\right)
^{-1/3}\right] \text{\ large }\frac{\Delta T}{T_{h}}g\overline{T} \\ 
\frac{4}{27}\left( \frac{\Delta T}{T_{h}}\right) ^{2}g\overline{T}\;\text{%
small }\frac{\Delta T}{T_{h}}g\overline{T}%
\end{array}%
\right.  \label{eta}
\end{equation}%
thus providing simple analytical expressions for these two regimes. The
small $\left( \Delta T/T_{h}\right) g\overline{T}$ expression provides a
good estimate of the full curve up to about 1\% efficiency, while the large $%
\left( \Delta T/T_{h}\right) g\overline{T}$ approximation is valid for $\eta
\gtrsim 20\%$.

\begin{figure}[h]
\centering \includegraphics[scale=0.6,clip=true]{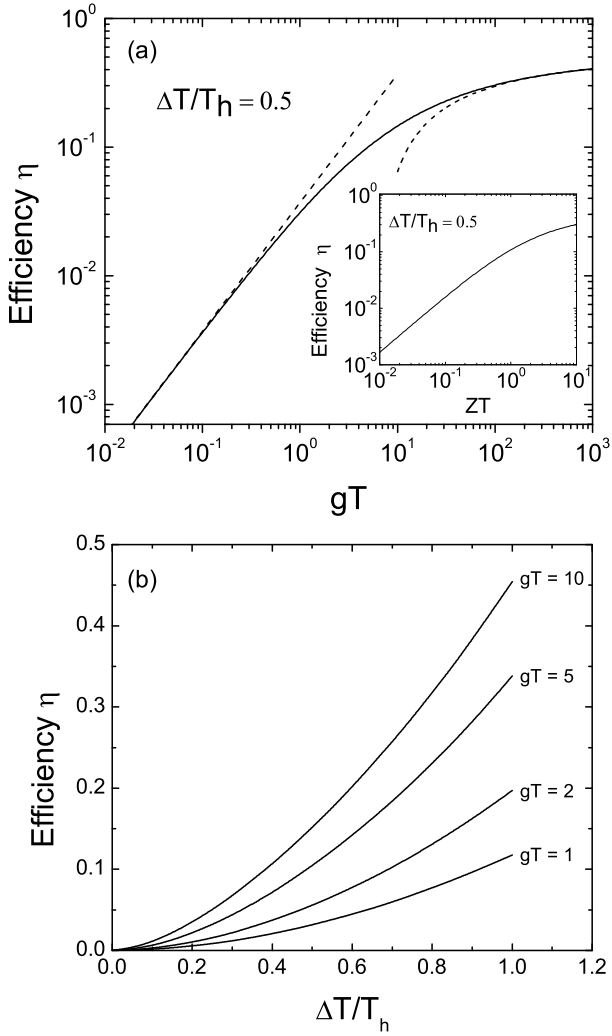}
\caption{(a) Thermoelectric efficiency as a function of the $g\overline{T}$
factor. Dashed lines are the low and high $g\overline{T}$ behaviors from Eq.
(\protect\ref{eta}). Inset is the efficiency as a function of $zT$. (b)
Thermoelectric efficiency as a function of $\Delta T/T_{h}$.}
\end{figure}

The inset in Fig. 3a shows the efficiency for a traditional ohmic material
as a function of $zT$ for $\Delta T/T_{h}=1/2$, where $zT=1$ gives about
10\% efficiency. Since this is the current state-of-the-art in
thermoelectric materials, we can use it as a comparison with the SCL regime;
in that case one would need $g\overline{T}\approx 5$ to achieve the same
efficiency at this given $\Delta T/T_{h}$. However, the efficiency depends
significantly on $\Delta T/T_{h}$, as shown in Fig. 3b for values of $g%
\overline{T}$ as high as 10. The curve for $g\overline{T}=1$ is well
approximated by the small $\left( \Delta T/T_{h}\right) g\overline{T}$ limit
of Eq. $\left( \ref{eta}\right) $, indicating a quadratic dependence on $%
\Delta T/T_{h}$.

The full expression for $g\overline{T}$ depends on the parameter $\xi $,
which in turn depends on the dimensionality of the material under
consideration; for bulk\cite{grinberg}, thin film\cite{grinberg}, and
nanowire\cite{talin} materials, it is given by%
\begin{equation}
\xi ^{2}=\left\{ 
\begin{array}{l}
\frac{9\varepsilon \mu }{8L^{3}}\;\text{bulk} \\ 
\xi _{0}\frac{\varepsilon \mu }{tL^{2}}\;\text{thin film, }t\ll L \\ 
\xi _{0}\frac{\varepsilon \mu }{R^{2}L}\;\text{nanowire, }R\ll L%
\end{array}%
\right.  \label{expansion}
\end{equation}%
where $\varepsilon $ is the permittivity, $\mu $ is the mobility, and $\xi
_{0}$ is a numerical constant that depends on the shape of the electrodes,
equal to 1 for planar contacts\cite{grinberg}. $t$ is the thickness of the
thin film, while $R$ is the nanowire radius (see Fig. 4 for illustrations of
the different geometries.) The expressions for the thin film and nanowire
cases are applicable when $t/L\ll 1$ and $R/L\ll 1$; for larger values of $%
t/L$ and $R/L$, $\xi $ crosses over to the bulk expression\cite{talin}.

The $g\overline{T}$ factor also depends on the material parameters $S$, $%
\kappa $, $\mu $, and $\varepsilon $. Like traditional ohmic materials, high 
$g\overline{T}$ requires large $S$, low $\kappa $ and large $\mu $; but it
also needs large $\varepsilon $, a criterion not usually required for high $%
zT$. This arises because large $\varepsilon $ serves to screen the injected
charge, and reduces the repulsive Coulomb interaction that opposes charge
injection. This is beneficial because dielectric screening maintains the SCL
transport regime, in contrast to free-carrier screening that would make the
current ohmic.

Two advantages of the SCL regime are that the maximum values of $S$ and $\mu 
$ can be exploited. Indeed, ohmic materials are usually operated at high
doping where both $S$ and $\mu $ are reduced from their maximum values. But
at low doping $S$ attains a maximum at a value\cite{goldsmid} $S\approx
E_{g}/2e\overline{T}$ while $\mu $ saturates to its intrinsic value $\mu
_{int}$. We can thus write the maximum value of $g\overline{T}$ as%
\begin{equation}
g\overline{T}=\left( \frac{E_{g}}{2e\overline{T}}\right) ^{3}\frac{%
a\varepsilon \mu _{int}LT_{h}^{2}}{\kappa }  \label{gT}
\end{equation}%
where $a$ is a geometry-dependent factor that can be obtained from Eq. $%
\left( \ref{expansion}\right) $. We note that the $g\overline{T}$ factor
depends linearly on $\mu $, just like an ohmic material where $zT\sim \sigma
\sim \mu $; however, the difference is that the mobility in $zT$ is the
high-doping mobility, which is smaller than $\mu _{int}$.

\section{Practical estimates}

To evaluate the practicality of the SCL approach for thermoelectric power
generation, we apply the theory to the bulk, thin film, and nanowire
geometries for $T_{h}=600$ K and $\Delta T$ = 300K. We consider GaAs at low
doping as an example material since SCL has been observed in this material%
\cite{allen}, and because material properties are readily available\cite%
{blakemore}: at $\overline{T}=$ $450$K and low doping, $E_{g}=$1.35 eV, $%
\varepsilon =13$, $\mu \approx 3150$ cm$^{2}/VS$, and $\kappa \approx $ 30
W/mK. Figure 4a shows the calculated efficiency for a bulk device as a
function of the channel length $L$. The efficiency is low unless $L\ $is
very small. The case of thin films (Fig. 4b) is slightly more promising with
larger efficiencies at longer lengths, but this requires small film
thicknesses. Also note that the large $\Delta T$ over the small lengths $L$
where reasonable efficiency is observed in Fig. 4a and 4b would lead to very
large heat fluxes that further limit the usefulness of these geometries.
Thus, for the bulk or thin film geometries to be useful, a much larger $g%
\overline{T}$ is needed so they can operate at longer lengths, and this
would require other or new materials with better properties for SCL
thermoelectrics.

\begin{figure}[h]
\centering \includegraphics[scale=0.6,clip=true]{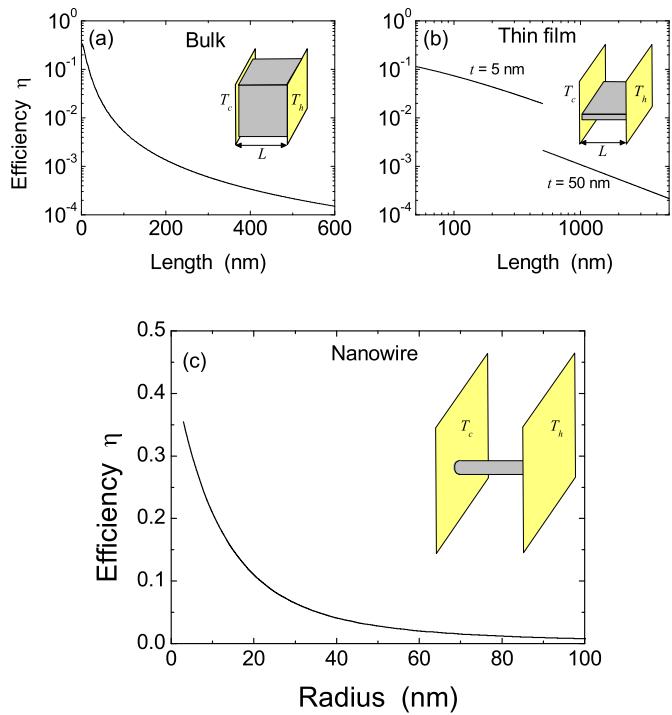}
\caption{Thermoelectric efficiency at $T_{h}=600K$ and $\Delta T=300K$ for
(a) bulk material as a function of length, (b) thin film material as a
function of length for two thicknesses, and (c) nanowire as a function of
radius. In (b), the curves are plotted for $t<L/10$ to satisfy the criterion 
$t\ll L$ (see Eq. \protect\ref{expansion}).}
\end{figure}

The situation is significantly more favorable if one considers nanowires, as
shown in Fig. 4c. In that case, SCL currents are enhanced due to the scaling 
$\xi ^{2}\sim R^{-2}L^{-1}$; in fact, $g\overline{T}$ becomes independent of
length, and scales as $R^{-2}$, with $R$ being naturally small for
nanowires. This leads to a large efficiency which exceeds the 10\% value for
nanowires less than 20nm in radius. This efficiency could be even larger if
other effects predicted and measured for nanowires were included; for
example, from Eq. $\left( \ref{gT}\right) $, $g\overline{T}$ depends
sensitively on the bandgap, which increases with decreasing diameter.
Similarly, the thermal conductivity is reduced with decreasing nanowire
diameter, which could also make $g\overline{T}$ larger. It should be noted
that the electrical currents needed to achieve the efficiency plotted in
Fig. 4c are less than a $\mu A$ per nanowire and require voltages on the
order of a volt, which is easily achieved without damaging the nanowires or
the contacts\cite{talin}.

\section{Conclusion}

In summary, we considered the efficiency of thermoelectric materials in the
space-charge-limited regime and found that it depends on a single
dimensionless parameter $g\overline{T}$, in analogy with conventional ohmic
materials. When applied to bulk, thin film and nanowire geometries, we find
that nanowires are the most promising to harness SCL transport. This work
provides a new path for improving the performance of thermoelectric
materials, and suggests the exploration of new thermoelectric materials with
properties conducive to SCL transport.

\section{Acknowledgement}

Discussions with Peter Sharma and Doug Medlin are gratefully acknowledged.
This project is supported by the Laboratory Directed Research and
Development program at Sandia National Laboratories, a multiprogram
laboratory managed and operated by Sandia Corporation, a wholly owned
subsidiary of Lockheed Martin Corporation, for the United States Department
of Energy's National Nuclear Security Administration under Contract
DE-AC04-94AL85000.

$^{\ast }$email:fleonar@sandia.gov


\begin{thebibliography}{99}
\bibitem{mahan} G. Mahan, J. Appl. Phys. {\bf 76}, 4362 (1994).

\bibitem{shakouri} A. Shakouri, Annu. Rev. Mater. Res. {\bf 41}, 399 (2011).

\bibitem{nolas} G. S. Nolas, J. Sharp, and H. J. Goldsmid, {\it %
Thermoelectrics Basic Principles and New Materials Developments }(Springer,
Berlin, 2001).

\bibitem{ulrich} M. D. Ulrich, P. A. Barnes, and C. B. Vining, J. Appl.
Phys. {\bf 90}, 1625 (2001).

\bibitem{mahan2} G. D. Mahan, J. O. Sofo, and M. Bartkowiak, J. Appl. Phys. 
{\bf 83}, 4683 (1998).

\bibitem{mahan3} G. D. Mahan, J. Appl. Phys. {\bf 87}, 7326 (2000).

\bibitem{zeng} T. Zeng and G. Xhen, J. Appl. Phys. {\bf 92}, 3152 (2002).

\bibitem{odwyer} M. F. O'Dwyer, T. E. Humphrey, R. A. Lewis, and C. Zhang,
J. Phys. D: Appl. Phys. {\bf 39}, 4153 (2006).

\bibitem{lampert} M. A. Lampert and P. Mark, {\it Charge Injection in Solids 
}(Academic Press, New York, 1970).

\bibitem{rose} A. Rose, Phys.\ Rev. {\bf 97}, 1538 (1955).

\bibitem{grinberg1} A. A. Grinberg and S. Luryi, J. Appl. Phys.{\bf \ 61},
1181 (1987).

\bibitem{davids} P. S. Davids, I. H. Campbell, and D. L. Smith, J. Appl.
Phys. {\bf 82}, 6319 (1997).

\bibitem{chandra} W. Chandra, L. K. Ang, and W. S. Koh, J. Phys. D: Appl.
Phys. {\bf 42}, 055504 (2009).

\bibitem{snyder} G. J. Snyder and T. S. Ursell, Phys. Rev. Lett. {\bf 91},
148301 (2003).

\bibitem{grinberg} A. A. Grinberg, S. Luryi, M. R. Pinto, and N. L. Schryer,
IEEE Trans. Electron Devices {\bf 36}, 1162 (1989).

\bibitem{leonard} The heat equation for nanowires is discussed in F. L\'{e}%
onard, Appl. Phys. Lett. {\bf 98}, 103101 (2011).

\bibitem{katzenmeyer} A. M. Katzenmeyer {\it et al}, IEEE Trans. Nanotech. 
{\bf 10}, 92 (2011).

\bibitem{harman} T. C. Harman and J. M. Honig, {\it Thermoelectric and
Thermomagnetic Effects and Applications} (McGraw-Hill, New York, 1967).

\bibitem{talin} A. A. Talin, F. L\'{e}onard, B. S. Swartzentruber, X. Wang,
and S. D. Hersee, Phys. Rev. Lett. {\bf 101}, 076802 (2008).

\bibitem{goldsmid} H. J. Goldsmid and J. W. Sharp, J. Electron. Mater. {\bf %
28}, 869 (1999).

\bibitem{allen} J. W. Allen and R. J. Cherry, Nature {\bf 189}, 297 (1961).

\bibitem{blakemore} J. S. Blakemore, J. Appl. Phys. {\bf 53}, R123 (1982).
\end{thebibliography}
\end{document}